\documentclass[12pt,twoside]{article}
\usepackage{xrb2007}
\usepackage{natbib,graphics,graphicx,subfigure}

\newcommand{\eg}{{\sl e.g.~}}

\newcommand{\etal}{{\sl et al.}}

\newcommand{\Chandra}{{\it Chandra~}}

\begin{document}

% select your session by uncommenting the appropriate line
%\session{Jets}
%\session{Jet and Black Hole Binaries}
%\session{Faint Galactic XRB Populations}
%\session{Faint XRBs and Galactic LMXBs}
%\session{Obscured XRBs and INTEGRAL Sources}
%\session{ULXs}
%\session{Extragalactic Populations}
\session{Future Missions and Surveys}
%\session{Population Synthesis}

\shortauthor{Raines, Eikenberry, and Bandyopadhyay}
\shorttitle{FLAMINGOS-2 Poster}

\title{FLAMINGOS-2: A Near-IR Multi-Object Spectrometer Ideal for Surveying the Galactic Center}
\vspace{-3mm}
\author{S.\ Nicholas Raines, Stephen S.\ Eikenberry, and Reba M.\ Bandyopadhyay}
\affil{University of Florida, Department of Astronomy, 211 Bryant Space Science Center, Gainesville, FL 32611}

\begin{abstract}
FLAMINGOS-2 (PI: S.\ Eikenberry) is a \$5M facility-class near-infrared 
(1-2.5 \micron) multi-object spectrometer and wide-field imager being built
at the University of Florida for Gemini South.  Here we highlight the 
capabilities of FLAMINGOS-2, as it will be an ideal instrument for surveying 
the accreting binary population in the Galactic Center.
\end{abstract}

\vspace{-7mm}
\section{Introduction}
\Chandra observations have shown that there is a much larger
population of accreting binaries in our Galaxy than was previously
recognized, with a particular concentration in the Galactic Center
(GC; Wang \etal\ 2002; Muno \etal\ 2003).  IR spectroscopy is the only
way to definitively identify the true stellar counterparts to these
X-ray sources (\eg Bandyopadhyay \etal\ 1999).  Due to the extreme
field crowding, to successfully find the true counterparts to the XRBs
in the GC we need to obtain spectra of $\sim$1000-1500 IR stars - a
nearly infeasible task to perform via traditional longslit
single-object spectroscopy.

FLAMINGOS-2 (PI: S.\ Eikenberry; Eikenberry \etal\ 2006) is a
facility-class near-IR (1-2.5 \micron) multi-object spectrometer (MOS)
and wide-field imager being built at the University of Florida for
Gemini South.  Here we highlight the capabilities of FLAMINGOS-2, as
it will be an ideal instrument for surveying the accreting binary
population in the GC.  Utilizing custom aperture masks in a 2$\times$6
arcminute$^{2}$ field-of-view (FOV), simultaneous multi-object
spectroscopy of up to $\sim$90 targets will be possible at resolving
power $\sim$1300 in the $H+K$ band.  With this resolution and FOV,
combined with the Gemini 8-m aperture, our team at UF will be able to
efficiently perform the first spectroscopic survey of this GC
population with high S/N to a limiting magnitude of $K\sim$17 during
our guaranteed time (\emph{cf.}\ Eikenberry \etal, Bandyopadhyay
\etal, these proceedings).

\section{FLAMINGOS-2 Optical Path \& Components}
FLAMINGOS-2 has a fully cryogenic optical train, illustrated in Fig.\
1, with 9 spherical refractive elements and two front-surface gold
flat mirrors.  The window and 7 of the lenses are single-crystalline
CaF$_2$; the other two lenses are made from Ohara SFTM-16.
Progressing from the top to the bottom of the figure, light first
passes through the window to a focus.  The MOS wheel lies at the
telescope focus and contains an imaging aperture and, most
importantly, a selection of custom aperture masks (``mosplates''); it
also caries 6 long slits. It is immediately followed by a selectable
baffle, called the Decker wheel, and the field lens.  The optical path
then is folded by the flat mirrors to the other two elements of the
collimator optics.  At this point the beam is collimated.  Two wheels
carrying a selection of filters for imaging and spectroscopy are
spaced on either side of the Lyot wheel.  FLAMINGOS-2 was designed for
operation with the telescope's \emph{f/16} beam but it also can accept
the $\sim$\emph{f/30} beam from the Gemini Multi-Conjugate Adaptive
Optics (MCAO) system, and the Lyot wheel carries pupil stops for both
modes of operation.  The final mechanism is a wheel carrying a
selection of grisms; it also includes a clear aperture for the imaging
mode of operation.  Light is then re-imaged onto the Hawaii-II array
by a 6-element camera lens assembly.

Standard near-IR $J$, $H$, and $Ks$ filters are installed in one of
the filter wheels; Gemini may additionally offer a $Y$-band filter
(0.97-1.07 \micron).  Two specialty spectroscopy filters are installed
in the other filter wheel, one covering the $J+H$ bandpasses, the
other covering $H+K$.  Three grisms reside in the grism wheel, two
with moderate resolving power, $R (= \lambda/\delta\lambda) \sim1300$,
and one grism with high resolving power, $R \sim 3300$.  The $R \sim
1300$ grisms are used in conjunction with the $J+H$ or $H+K$ bandpass
filters, while the $R \sim 3300$ grism is used with the $J$, $H$, or
$Ks$ standard near-IR filters for out-of-bandpass blocking.

\begin{figure}[t!]
\centering
\includegraphics{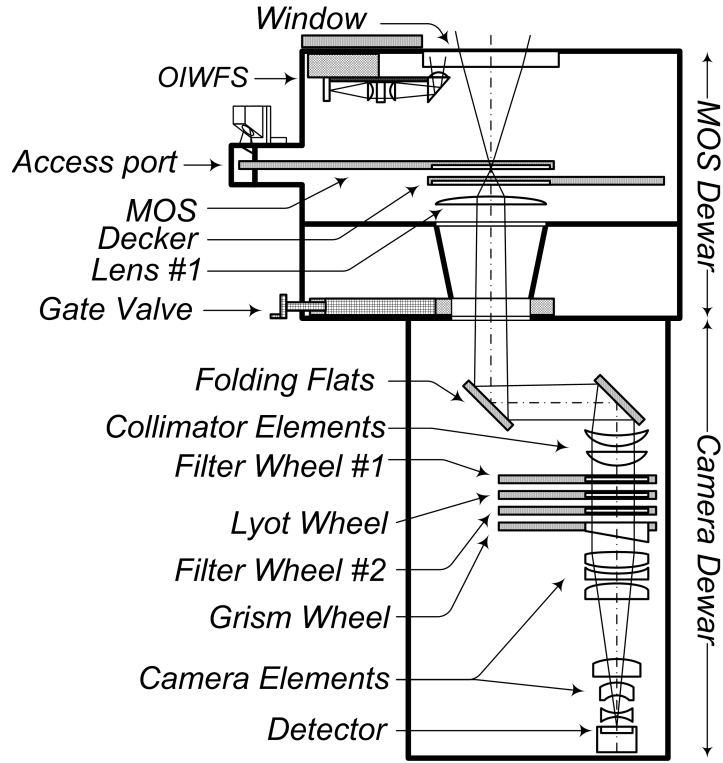}
\caption{Optical Path Diagram for FLAMINGOS-2}
\end{figure}

\vspace{-3mm}
\section{FLAMINGOS-2 MOS Mode }
Several features of FLAMINGOS-2 make it ideal for surveying the
Galactic Center: \emph{(a)} a wide imaging FOV of $\sim$6.2
arcminutes$^{2}$ ($\sim$3.1 arcminutes$^{2}$ with MCAO), \emph{(b)}
the ability to carry up to 9 custom mosplates at a time, \emph{(c)}
the mosplates' large spectroscopic FOV of 2$\times$6.2 square
arcminutes (1$\times$3.1 square arcminutes with MCAO), \emph{(d)} the
cooling of the masks to cryogenic temperatures which allow low
internal instrument background for operation in the K-band, and
\emph{(e)} the ability to quickly exchange the set of mosplates.

The MOS wheel, shown in Fig.\ 2a, is 0.9 meters in diameter. It has
three circular apertures positioned around the periphery of the wheel;
one is left open for imaging and the other two usually contain pinhole
masks for engineering.  Equally spaced between the circular apertures
are 9 rectangular slots for holding custom mosplates.  A test mosplate
is shown in Fig.\ 2b.  The mosplate FOV (\emph{f/16-mode}) has
sufficient sky-coverage to design custom mosplates containing up to
$\sim$90 slitlets.

Mosplates can be changed during the daytime without thermally cycling the
entire instrument.  Also shown in Fig.\ 1 is a \emph{gate valve}, positioned 
between the field lens and the folding flat mirrors.  At the end of a night of
 observing this valve is closed, the MOS dewar cooling is halted, and a 
warm-up heater is turned on.  Several hours later, during the daytime, an 
engineer can open the access port on the side of the MOS dewar.  
Each mosplate is held in a frame which slides into the edge of the MOS wheel.
The observed plates are removed, and a new set of mosplates, each one 
already mounted in a frame, are slid into place.  The engineer then closes up
the access port and begins the process of evacuating and cooling the dewar.
Once it is cold enough the gate valve is then re-opened.  
By design, this should be completed in time for the observers who return that
evening.

If each plate is observed for only 1 hour, $\sim$800 spectra could
be obtained with a single night's observation using all 9 mosplates.
With only three nights of observing potentially up to $\sim$2400 spectra 
could be obtained.  Thus FLAMINGOS-2's MOS mode of operation is ideally suited
for identifying the true stellar counterparts to the X-ray sources in the 
GC.

\begin{figure}[t!]
\centering
\subfigure[F2 mos wheel]
{
  \label{fig:sub:a}
  \includegraphics{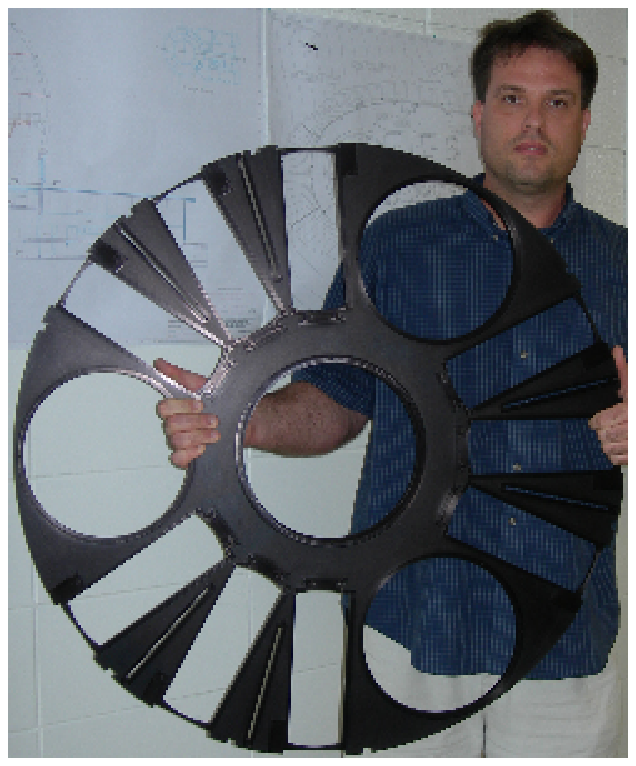}
}
\hspace{1cm}
\subfigure[Test Mosplate]
{
  \label{fig:sub:b}
  \includegraphics{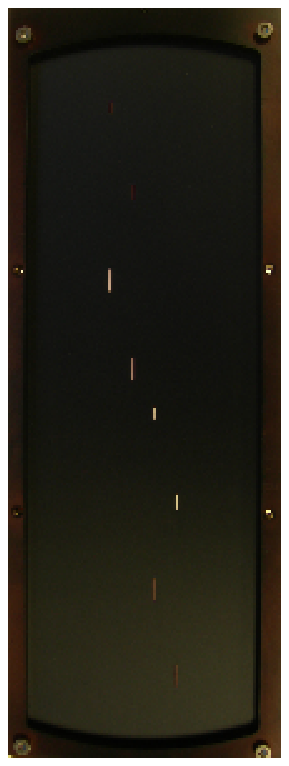}
}
\caption{(a) Engineer holding the 0.9 m diameter moswheel. One of the three circular apertures around the periphery is for imaging and is $\sim$234 mm in diameter, which corresponds to $\sim$6.28 arcmin. The wheel can carry 9 mosplates. (b) A FLAMINGOS-2 test mosplate; 8 slitlets of varying lengths are visible. The region interior to the mounting frame is $\sim$76 mm x $\sim$232 mm, which corresponds to $\sim$2.0 arcmin x $\sim$6.2 arcmin.}
\end{figure}

\end{document}